\documentclass[manuscript]{acmart}

\AtBeginDocument{%
  \providecommand\BibTeX{{%
    \normalfont B\kern-0.5em{\scshape i\kern-0.25em b}\kern-0.8em\TeX}}}

\copyrightyear{2023}
\acmYear{2023}
\setcopyright{acmlicensed}\acmConference[CHI '23]{Proceedings of the 2023
CHI Conference on Human Factors in Computing Systems}{April 23--28,
2023}{Hamburg, Germany}
\acmBooktitle{Proceedings of the 2023 CHI Conference on Human Factors in
Computing Systems (CHI '23), April 23--28, 2023, Hamburg, Germany}
\acmPrice{15.00}
\acmDOI{10.1145/3544548.3581405}
\acmISBN{978-1-4503-9421-5/23/04}

\usepackage{multirow}
\usepackage{colortbl}
\usepackage{makecell}
\usepackage{csquotes}
\usepackage{array}
\usepackage{CJKutf8}
\usepackage{graphicx}
\usepackage{steinmetz}
\newcommand{\PreserveBackslash}[1]{\let\temp=\\#1\let\\=\temp}

\newcolumntype{C}[1]{>{\PreserveBackslash\centering}p{#1}}
\newcolumntype{R}[1]{>{\PreserveBackslash\raggedleft}p{#1}}
\newcolumntype{L}[1]{>{\PreserveBackslash\raggedright}p{#1}}

\def\markup{0}

\if\markup1
\newcommand{\rv}[1]{{\leavevmode\color{blue}#1}}
\else
\newcommand{\rv}[1]{#1}
\newcommand{\st}[1]{}
\newcommand{\sout}[1]{}
\fi

\begin{document}

\title[Exploring How VR Supports Remote Communication Between Grandparents and Grandchildren]{Bridging the Generational Gap: Exploring How Virtual Reality Supports Remote Communication Between Grandparents and Grandchildren}


\author{Xiaoying Wei}
\orcid{0000-0003-3837-2638}
\affiliation{
  \institution{The Hong Kong University of Science and Technology}
  \city{Hong Kong SAR}
  \country{China}
}
\email{xweias@connect.ust.hk}

\author{Yizheng Gu}
\orcid{0000-0002-1794-2171}
\affiliation{
  \institution{Tsinghua University}
  \city{Beijing}
  \country{China}
}
\email{zsyzgu@163.com}

\author{Emily Kuang}
\orcid{0000-0003-4635-0703}
\affiliation{
  \institution{Rochester Institute of Technology}
  \city{Rochester}
  \country{USA}
}
\email{ek8093@rit.edu}

\author{Xian Wang}
\orcid{0000-0003-1023-636X}
\affiliation{
  \institution{The Hong Kong University of Science and Technology}
  \city{Hong Kong SAR}
  \country{China}
}
\email{xian.wang@connect.ust.hk}

\author{Beiyan Cao}
\orcid{0000-0001-6524-7985}
\affiliation{
  \institution{The Hong Kong University of Science and Technology}
  \city{Hong Kong SAR}
  \country{China}
}
\email{beiyan.cao@connect.ust.hk}

\author{Xiaofu Jin}
\orcid{0000-0003-3837-2638}
\affiliation{
  \institution{The Hong Kong University of Science and Technology}
  \city{Hong Kong SAR}
  \country{China}
}
\email{xjinao@connect.ust.hk}

\author{Mingming Fan}
\authornote{Corresponding Author}
\orcid{0000-0002-0356-4712}
\affiliation{
  \institution{The Hong Kong University of Science and Technology (Guangzhou)}
  \city{Guangzhou}
  \country{China}
}
\affiliation{
  \institution{The Hong Kong University of Science and Technology}
  \city{Hong Kong SAR}
  \country{China}
}
\email{mingmingfan@ust.hk}

\renewcommand{\shortauthors}{Xiaoying Wei, et al.}

\begin{abstract}

When living apart, grandparents and grandchildren often use audio-visual communication approaches to stay connected. However, these approaches seldom provide sufficient companionship and intimacy due to a lack of co-presence and spatial interaction, which can be fulfilled by immersive virtual reality (VR). To understand how grandparents and grandchildren might leverage VR to facilitate their remote communication and better inform future design, we conducted a user-centered participatory design study with twelve pairs of grandparents and grandchildren. Results show that VR affords casual and equal communication by reducing the generational gap, and promotes conversation by offering shared activities as bridges for connection. Participants preferred resemblant appearances on avatars for conveying well-being but created ideal selves for gaining playfulness.
Based on the results, we contribute eight design implications that inform future VR-based grandparent-grandchild communications.

\end{abstract}





\maketitle

\section{Introduction}

Grandparents and grandchildren can benefit from close intergenerational communication. By engaging in close inter-generational communication, grandchildren can gain a stronger sense of connection with the larger family and foster their self-development (e.g., emotional development and self-esteem)~\cite{brussoni1998grandparental, Tomlin1998handbook, williams1997young}. In return, grandparents gain stronger social interactions and feel less socially isolated~\cite{lindley2009desiring, cornejo2013enriching, seifert2020digital, nicholson2012review, cotterell2018preventing}. However, today's families are becoming increasingly dispersed due to many factors, including the younger generation relocating for school or work~\cite{stafford2004maintaining, vutborg2010family}, and this imposes challenges to maintaining close connections between grandparents and grandchildren. 
 \rv{Although audio-visual media (e.g., phone calls, video calls, or instant messaging) allow remote family members to see or hear each other~\cite{judge2010sharing}, they are limited in providing a sufficient sense of togetherness and companion~\cite{tu2002relationship, biocca2003toward}. Moreover, these media do not offer grandparents and grandchildren something to do together other than talking, which makes it hard to sustain their attention during communication and may decrease their conversation quality~\cite{raffle2011hello, viguer2010grandparent, richter2018relations, mcauley2012spending, fuchsberger2021grandparents}. }
 \sout{Although common audio-visual communication approaches (e.g., videoconferencing (VC) and phone calls) are used to maintain remote inter-generational relationships~\cite{harwood2000communication, fuchsberger2021grandparents}, they have two shortcomings in terms of supporting inter-generational communication: (1) they fall short in providing a sufficient sense of \textbf{immersion and social presence}, thus making it difficult to maintain physical and emotional closeness with their remote relatives~\cite{tu2002relationship, biocca2003toward}; (2) they offer no or limited \textbf{spatial interaction} opportunities compared to face-to-face (F2F) communication, such as walking to a garden while chatting~\cite{fuchsberger2021grandparents}. With the absence of such spatial interactions, it is less convenient for grandparents and grandchildren to share their thoughts and emotions, which reduces the enjoyment gained from their communication~\cite{viguer2010grandparent, richter2018relations}. }

 Virtual reality has the potential to mitigate these limitations of traditional audio-visual media in two ways. First, users communicate and interact via embodied avatars---digital representations of a human user. Second, users can interact with each other within a shared 3D virtual world. These attributes of VR show potential advantages over traditional audio-visual media. First, communicating via avatars in a virtual world allows users to perceive a better sense of co-present and togetherness~\cite{Li2019, smith2018communication}, collapsing the physical and emotional distance often experienced when using other communication media~\cite{Tanenbaum2020, bailenson2004transformed}. Second, users can ``physically'' interact with each other via avatar as in person, which enables them to express themselves more freely~\cite{smith2018communication, abdullah2021videoconference}. Lastly, users could participate in varied shared activities in VR, such as doing sports~\cite{hoeg2021buddy}, sharing photos~\cite{Li2019}, and learning together~\cite{wang2022construction}.
 While prior work showed that older adults and their younger generations feel enjoyable and understand each other better by doing shared VR activities (e.g., tandem biking, co-learning) in co-located scenarios~\cite{wang2022construction, hoeg2021buddy}, only one recent work, to our knowledge, explored VR as a communication medium for older adults and their family members who live at a distance~\cite{afifi2022using}. \rv{While this paper suggested that VR-based remote communication can improve the overall life quality of older adults with cognitive impairments and reduce the burden of remote family members~\cite{afifi2022using}, it remains unknown how grandparents and grandchildren would want VR to be designed to facilitate their remote inter-generational communication.}
  \sout{Prior research has explored communication experiences in VR among different groups of people, and showed that they benefit from different advantages brought by VR (e.g., children prefer rich social interactions~\cite{maloney2021stay}, older adults enjoy the sense of anonymity~\cite{BakerJ2021}, and long-distance couples are pleased by embodied physical contact in VR~\cite{zamanifard2019togetherness}).}
 \sout{However, no prior work investigated how VR can be leveraged to promote inter-generational communication between grandparents and grandchildren. Although recent research has showed that remote communication in VR can improve the overall life quality of older adults and their remote family~\cite{afifi2022using}, it did not examine what factors in VR contribute to their communication experiences and how to design better VR-based remote inter-generational communication in this context. }
 
  
  \rv{Inspired by this line of work, we take a step further to answer the following research question (RQ): \textit{How might VR be leveraged to facilitate inter-generational communication between grandparents and grandchildren?} Specifically, how would they want to be represented by avatars? What types of interpersonal interactions do they expect? What types of shared activities would they want to engage in? }
 To answer these questions, we conducted a user-centered participatory design study with twelve pairs (N=24) of grandparents and grandchildren. \rv{We investigated participants' preferences and visions in three key aspects of VR: avatar appearance, interpersonal interaction, and shared activity, which reflect how they see themselves, each other, and the environment in which they interact. We also explored their perceived barriers to adopting VR as their communication tool.}

 
 Our results show that when communicating in VR, \rv{1) grandparents and grandchildren chose different appearances of avatars to meet different communication needs: they chose avatars that reflect their current physical appearances to check up on each other's well-being, while preferring avatars reflecting the ideal versions of themselves (e.g., younger/better-looking) to gain better communication experience (e.g., confident, reminiscing); 2) VR affords casual and equal communication between grandparents and grandchildren because it could make ``intimate'' or ``inappropriate'' interpersonal interactions between them, that they hesitate to do in real life, more acceptable and welcoming. Moreover, sensory feedback beyond vision and audio may help enhance their feelings of togetherness and alleviate their feelings of nostalgia when living apart; 3) participants felt that engaging in shared activities in VR allowed them to find more topics to fill in the silence in conversations and keep their conversations flowing. They further envisioned three types of shared activities they wanted to engage in, including exploring difficult-to-visit locations, reminiscing together, and re-experiencing daily activities; 4) the overhead of setting up VR for communication, the perceived difficulty of certain VR operations, and safety concerns negatively impact participants' willingness to use VR for inter-generational communication.}
 Building on top of our findings, we contribute eight design implications for VR researchers and developers to consider when leveraging VR to support remote communication between grandparents and grandchildren.
 In sum, we make the following contributions: 
 \begin{itemize}
    \item We identified grandparents' and grandchildren's preferences and visions for avatar appearance, interpersonal interaction, and shared activities to better support them in communicating with each other in VR.
    \item We derived and discussed design implications for designing VR to better facilitate intergenerational communication between grandparents and grandchildren.
\end{itemize}

\section{Related Work}

\subsection{Inter-generational Remote Communication}

 Both grandparents and grandchildren can benefit from a positive inter-generational connection~\cite{brussoni1998grandparental, Tomlin1998handbook, harwood2000communication, harwood2000affiliation, lindley2009desiring}. Grandparents and grandchildren view each other as windows into their corresponding times and treasure the values and experiences learned from each other~\cite{seponski2009caring, kornhaber1981grandparents}. However, this relationship is challenging to maintain due to the geographic distance, which is commonly caused by the migration of the younger generation for study or work~\cite{stafford2004maintaining, vutborg2010family}. The widespread COVID-19 pandemic and its impacts on travel and quarantines also aggravated this situation~\cite{heshmat2021family}. 
 
 People have a strong need and desire to stay connected with remote families to share their recent life and well-being \rv{(e.g., their mental and physical health~\cite{dunn1959high}})~\cite{mynatt2001digital, tee2009exploring, romero2007connecting, neustaedter2006interpersonal}. 
\sout{Currently, they use audio-visual communication methods (e.g., phone calls, video calls, or instant messaging) to stay in contact with remote family members~\cite{cao2010understanding, ballagas2009family, raffle2011hello, wen2020coronavirus, neustaedter2006interpersonal, tee2009exploring}. A decade ago, the phone was considered the most vital communication tool between grandchildren and grandparents~\cite{ballagas2009family}. Nowadays, the most dominant form of communication is videoconferencing, where users can see each other in addition to hearing their voices~\cite{fuchsberger2021grandparents}.}
 Currently, grandparents and grandchildren use audio-visual communication methods to stay in contact with remote family members~\cite{cao2010understanding, ballagas2009family, raffle2011hello, wen2020coronavirus, neustaedter2006interpersonal, tee2009exploring}. However, these media are limited in catering to their need for getting a better sense of togetherness and companion~\cite{tu2002relationship, biocca2003toward}. Additionally, grandparents and grandchildren hope to communicate with each other while having something to do to foster communication and enjoy the time they spend together~\cite{raffle2011hello, viguer2010grandparent, richter2018relations, mcauley2012spending, fuchsberger2021grandparents}, which current audio-visual communication platforms can not provide. 

 Inspired by recent work that suggested VR has the potential to overcome these limitations by allowing users to communicate and interact in an immersive virtual environment with embodied avatar~\cite{Li2019, afifi2022using, tham2018understanding}, we sought to understand how grandparents and grandchildren would like to leverage VR to facilitate their remote communication.

 \subsection{VR as a Communication Tool}
 \sout{With VR, remote users can enter a shared virtual environment to communicate with others in novel and immersive ways~\cite{tham2018understanding, Li2019}. Previous studies focused on using VR as a communication tool to facilitate multi-user collaboration or co-design activities. For instance, VR was employed to enhance design-related activities in engineering companies with remote colleagues~\cite{berg2017industry}. Zimmerman's survey showed that VR helps cross-disciplinary communication by providing scenarios for the designer and product manager to communicate together~\cite{zimmermann2008virtual}. Mei et al. proposed a VR platform that allows clients to remotely co-design cakes with chefs through 3D visualizations~\cite{mei2021cakevr}. In recent years, various social VR platforms have increasingly flourished on the commercial market (e.g., Rec Room, BigScreen, AltspaceVR, and VRChat), which allows people to meet, interact, and socialize in more appealing ways, such as throwing parties, watching movies, and attending events and performances while being represented by avatars~\cite{freeman2022working, mcveigh2018s}. Researchers have also focused on the social aspects of VR and explored how users can benefit from communication in VR~\cite{Li2019, abdullah2021videoconference}. }
 
 \sout{VR brings several advantages for communication. Maloney et al. explored various non-verbal communication strategies in VR and suggested that it can positively affect users' communication experiences (e.g., offering immersive and embodied interactions, privacy, and social comfort)~\cite{Maloney2020}. Smith et al. suggested that VR provided a high level of co-presence~\cite{smith2018communication}, which combines the benefits of traditional communication media and physical interactions while providing an easier way to initiate social connection~\cite{Tanenbaum2020, smith2018communication, bailenson2004transformed}.}
 
 \rv{As social VR platforms (e.g., VRChat, Rec Room, BigScreen, and AltspaceVR) increasingly emerged on the market, researchers began to investigate how people interact with each other in social VR platforms~\cite{tham2018understanding, freeman2022working, mcveigh2018s}.}
 \rv{Research shows that compared to communicating through traditional visual-audio channels~\cite{wei2022communication}, people feel a higher level of co-presence with others~\cite{smith2018communication, Li2019}, are able to perform embodied physical interactions~\cite{zamanifard2019togetherness} and engage in varied activities with others in VR~\cite{freeman2022working, Li2019}. Moreover, VR could provide an easier way to initiate social interaction~\cite{Tanenbaum2020, smith2018communication, bailenson2004transformed}. }

 In addition to the general users, researchers began to explore how VR might be used by older adults to maintain their social connections~\cite{baker2021avatar, baker2019exploring, baker2021school, afifi2022using, hoeg2021buddy}. For example, Baker and his colleagues reported a series of studies investigating how older adults communicate with each other in VR, including their perception of avatar-media communication~\cite{baker2021avatar} and their favorite shared activities with friends~\cite{baker2019exploring, baker2021school}. 
 \rv{Recently, researchers also investigated how VR might be used for older adults and the younger generation. By engaging in specific VR activities (e.g., tandem biking, co-learning)}, older adults and their younger generations could gain positive emotions and understand each other better afterward~\cite{wang2022construction, hoeg2021buddy}. Communication in VR can also improve the overall quality of life for older people with cognitive impairments while decreasing the caregiver burden of their younger generations who live at a distance~\cite{afifi2022using}.
 While informative, prior research mostly used predefined activities to investigate how older adults communicate with other older adults or younger generations in VR. However, it remains unknown how best to design VR to facilitate inter-generational communication between  grandparents and grandchildren. 
 In this paper, we explored grandparents' and grandchildren's VR preferences and visions regarding \textit{avatar appearance}, \textit{interpersonal interaction}, and \textit{shared activities}. We also explored their perceived barriers to adopting VR as a communication tool.

 \st{In this paper, we identify the pros and cons of VR in remote inter-generational communication and explore grandparents' and grandchildren's preferences for VR-based remote communication. Based on these findings, we propose several design implications for how future VR might be used to facilitate communication between grandparents and grandchildren.}

 \subsubsection{Avatar Appearance}
 Avatar plays an essential role in VR by shaping users' communication experiences~\cite{garau2001impact, inkpen2011me}. Individuals often rely on the representation by avatars to symbolize themselves in social VR interaction~\cite{nowak2018avatars, manninen2007value, smith2018communication, freeman2021body}. People view avatars as their embodied interface~\cite{waltemate2018impact} and the \textit{protective shield}~\cite{bosch2013professional} in VR. When represented by avatars, they perceive a sense of relaxation and safety in conversations~\cite{BakerJ2021, Maloney2020, abdullah2021videoconference, Ide2020, bosch2013professional}, and feel closer to the others physically and emotionally than communicating via videoconferencing.  ~\cite{Li2019, Maloney2020, maloney2021stay}.
 
 Previous research also explored avatar creation and appearance~\cite{lin2014avatar, banakou2018virtually, kolesnichenko2019understanding, campbell1992body} and found that people tend to create their avatars meticulously. They also have different motivations for creating their avatar (e.g., idealized self, standing out, following a trend)~\cite{ducheneaut2009body, lin2014avatar} and often change their avatar's appearance to suit the specific communication context~\cite{baker2021avatar, kafai2010your}. Research on avatar appearances in VR suggested that it may affect not only users' own experiences (e.g., gaining more confidence~\cite{maister2013experiencing, banakou2018virtually}) but also others' perceptions of them (e.g., the attractiveness of avatar appearances were positively correlated with friendliness)~\cite{latoschik2017effect}. For example, represented by avatars' appearance that is different from themselves, introverted older adults feel a higher level of anonymity and are more willing to engage in conversations with strangers, which may potentially reduce the risk of social isolation \cite{baker2021avatar}.
 While these studies provided insights into the design and effects of avatars, none focused on how grandparents and grandchildren would want themselves to be represented in VR when communicating with each other and how they would use their avatars to interact with each other. 

 \subsubsection{Interpersonal interaction}
 \rv{Embodied avatars in VR enable different kinds of interpersonal interactions (e.g., hugs, pats) and new interaction experiences compared to the traditional communication platforms~\cite{maloney2021stay, freeman2022disturbing}. For instance, unsolicited touching from strangers is considered harassment~\cite{freeman2022disturbing}. People would feel offended when the interpersonal distance between them and strangers is too close~\cite{williamson2021proxemics, mueller2014proxemics}. 
 Previous research shows that different groups of people enjoy different kinds of interpersonal interaction in VR. Children are interested in simulated touch and social interactions that enable them to build a social connection with others~\cite{maloney2021stay}, and long-distance couples benefit from embodied physical contact and the sense of togetherness in VR~\cite{zamanifard2019togetherness}. 
 However, it remains unknown how grandparents and grandchildren would like to interact with each other via embodied avatars in VR. }

 \subsubsection{Shared Activities}
 In-person shared activities (e.g., collaboration, entertainment, or helping each other) are essential factors to enrich individuals' social lives~\cite{freeman2022working, reissman1993shared}. Engaging in these activities can foster the exchange of stories and experiences between grandparents and grandchildren, which benefits the self-development of grandchildren and increases the social connection of grandparents~\cite{viguer2010grandparent, mcauley2012spending}. Kennedy et al. studied the in-person activities of grandparents and grandchildren and clustered them into five categories: sociability, companionship, helping grandparent, community events, and grandparent entertaining~\cite{kennedy1992shared}.
 
 While such shared activities are often missing in video-based communication, VR could potentially enable users to participate in such activities through their embodied avatars in an immersive virtual world. 
 Previous research explored shared activities in VR in different populations, including teenagers, older adults, colleagues, and long-distance couples~\cite{maloney2020virtual, zamanifard2019togetherness, maloney2021stay, baker2019interrogating}, and showed that the preferences and unique needs for shared activities vary for different groups of people. 
 For instance, long-distance couples preferred to engage in real-life activities in VR~\cite{zamanifard2019togetherness} while older adults were more interested in reminiscing their prior experiences together and traveling~\cite{BakerJ2021, baker2019exploring, thach2020older}. 
 \rv{However, it remains unknown how grandparents and grandchildren would perceive the benefits of engaging in shared activities in VR and what types of shared activities that they would like to engage in. }
  Motivated by the literature and the gaps in the three aspects mentioned above, we took a step further to investigate these gaps through a participatory design study.

 \subsubsection{\rv{Barriers to adopting VR in older adults}}
 \rv{Prior work has examined how VR can be utilized to support healthy aging, including the use of VR as a tool to aid physical and cognitive health \cite{kim2017walking, garcia2015using, huang2020exergaming}, improving well-being and addressing depressive symptoms by using VR for enrichment \cite{thach2020older, baker2020evaluating, munusturalar2016virtual, houwelingen2021virtual}. On the one hand, older adults face certain barriers use VR, including its weight, tightness, the difficulty of operation, and causing feelings of being trapped, afraid, or anxious~\cite{healy2022older, roberts2019older}.
 On the other hand, older adults show a positive attitude toward VR technology as it could bring them opportunities to take part in new experiences and enrich their lives~\cite{riaz2021virtual, huygelier2019acceptance, arlati2021acceptance}. Previous work has explored how to help older adults to gain a better VR-using experience by getting support from caregivers in residential aged care houses or proposing guidelines for designing friendly VR applications.
 Inspired by prior work, we would also want to understand, in the context of inter-generational communication, the perceived barriers among grandparents and grandchildren when they communicate with each other in VR so that they could better leverage VR for inter-generational communication.
 }

\section{METHOD}

\begin{figure*}[]
\includegraphics[width=0.95\linewidth]{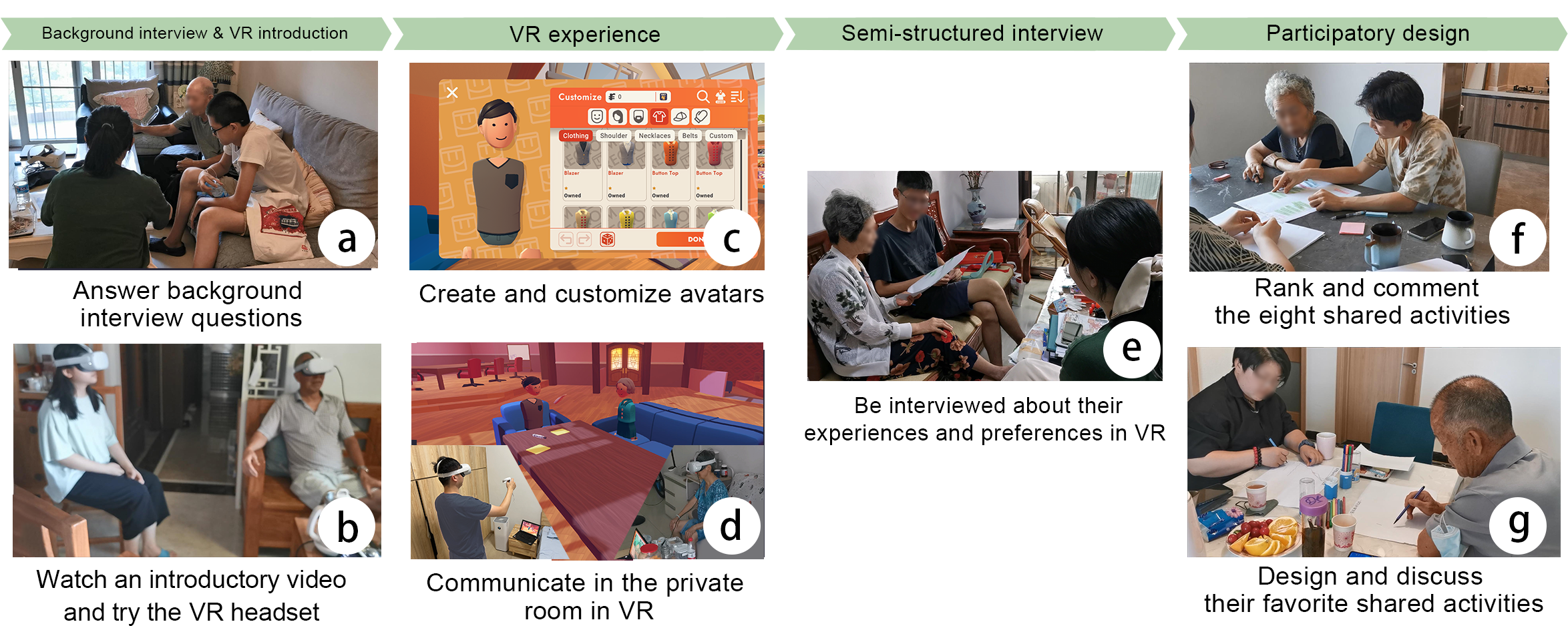}
\vspace{-0.2cm}
\caption{User study process where each column corresponds to one part of the study.} 
\Description{Photos taken during study sessions arranged in four columns where each column corresponds to each part of the study: (1) Background interview \& VR application, (2) VR experience, (3) Semi-structured interview, and (4) Participatory design.}
\label{fig:studyprocess}
\end{figure*}
 
We conducted a user-centered participatory design study with four parts as shown in Figure ~\ref{fig:studyprocess}: (i) background interview and introduction to VR; (ii) VR experience session in which grandparents and grandchildren talked and interacted with each other in the virtual environment; (iii) semi-structured interview for getting feedback regarding their communication experiences and preferences of avatar appearances in VR; (iv) participatory design session to elicit and understand their visions of VR shared activities. \st{The whole study lasted for about two and a half hours} \rv{The average length of experiments was two and a half hours}.

\subsection{Participants}

\st{To understand the needs of grandparents and grandchildren for remote communication,} We recruited grandparents and grandchildren \rv{who had lived separately for more than one year and did not currently live together}. We advertised our recruitment information through snowball sampling, social media, and community centers in three cities. Before the study, participants completed an online recruitment survey. Based on the responses, we invited participants with diverse backgrounds (e.g., education level, job) to participate in the full study.

 We recruited 24 participants (12 pairs of grandparents and grandchildren) from three locations (six pairs were from the metropolitan center, while six were from two regional locations 90 km from the metropolitan participants). The grandchildren were 15 to 27 years old ($M=21, SD=4$), with six females and six males. All the grandchildren have used VR at least once. The grandparents were 66 to 86 years old ($M=76, SD=4$), with six females and six males. Two older participants reported having difficulty walking. Each participant was compensated \$30 for their time. Table \ref{tab:participantInfo} shows participants' detailed backgrounds. \sout{We use F as shorthand for family, GP as shorthand for a grandparent, and GC as shorthand for a grandchild in the following sections.} In the paper, we use the following abbreviations: F (family), GF(grandfather), GM(grandmother), GS(grandson), and GD(granddaughter) in the following sections. For example, (F1, GM) means the grandmother in family 1.

\begin{table*}[!hbt]
\renewcommand{\arraystretch}{1.1}
    \caption{Participants' demographic information.}
    \label{tab:participantInfo}
    \begin{tabular}{|p{3.5cm}|p{4.5cm}|p{4.5cm}|}
    \hline
        & \textbf{Grandparent (12)}  & \textbf{Grandchildren (12)}          \\ \hline
        \textbf{Gender}  & Female (6), Male (6) & Female (6), Male (6)   \\ \hline
        \textbf{Age}     & 65-69(1), 70-74(1), 75-79(9), 85-89(1) & 15-19(4), 20-24(6), 25-29(2)   \\ \hline
        \textbf{Time Spent Using Smart Device per Day} & less than one hour (3), one to two hours (3), two to four hours (5), more than four hours (1) & less than one hour (1), one to two hours (2), two to four hours (4), more than four hours (5)  \\ \hline

        \textbf{On-line Communication Frequency} & \multicolumn{2}{c|}{\makecell[c]{less than once a month (0), about once a month (4), about twice a month (2), \\ once a week (3), more often(3)}}  \\ \hline
        
        \textbf{In-person Communication Frequency}     & \multicolumn{2}{c|}{\makecell[c]{less than once a year (1), about once a year (3), about twice a year (2), \\ about once in two months (1), about once a month (3), more often (2)}} \\   \hline
        
    \end{tabular}
\end{table*}

\subsection{Apparatus and Settings}
 
\textbf{Devices.} We used Oculus Quest 2~\cite{quest2Website} as the apparatus. Participants interacted with the VR system using two hand controllers. We set up two cameras in the experimental area to record their physical actions in two rooms, and the head-mounted display (HMD) projection on the PC's screen.

\textbf{Platform.} We chose Rec Room~\cite{recroomWebsite} as our communication platform in experiment. Rec Room allows participants to create and customize their avatars easily and enables participants to interact with players, as well as to provide various facilities (e.g., table tennis, poker, darts, and singing stage) for entertainment. \rv{This well-designed commercial social VR platform can be a vehicle for us to understand the challenge and needs of the communication between grandparents and grandchildren.}
Figure \ref{fig:recroom} shows some example scenes. 

\textbf{Choice of Shared Activities in VR.} In our participatory design session, we provided eight types of shared activities in VR for participants to rank. To select these preset activities, we first conducted pilot interviews to ask what shared activities in VR grandchildren most wanted to engage in with their grandparents with six participants (aged 18 - 27) who did not participate in the final user study. In the meantime, we reviewed previous work on activities that are most attractive to older adults~\cite{thach2020older, baker2019exploring}. The researchers then discussed and consolidated the results into a final list of eight activities based on the commonality of activities. The final results are shown in Table~\ref{fig:activities}.
 
 \subsection{Procedure}
 \st{As it was inconvenient for older adults to travel to our laboratory, we conducted all study sessions at the participants' homes. The familiar environments allowed participants to express their ideas more naturally. } Figure \ref{fig:studyprocess} shows each part of the user study.
 \rv{The studies were conducted at grandparent participants' homes, with their permission. Their grandchildren participants and the experimenters came to their homes to conduct the studies face-to-face. After the introduction of our study, with the help of experimenters, each pair of grandparent and grandchild were guided into two separate rooms in their home to experience VR communication to simulate remote conditions. Afterward, they returned to the same room to be interviewed and carry out the co-design part face-to-face in order to facilitate their discussion and co-creation process.}
 

 \subsubsection{Part One: Background Interview and VR introduction}
 In this session, we gave an overview of the session and conducted a background interview about their closeness, online/in-person meeting frequency, and favorite shared activities in person. Then participants watched a 3-minute video and used an application to get familiar with VR. Both the video and the VR application were created by the researchers. The video contained introductions to several types of popular VR applications, including traveling, watching movies, shooting games, role-playing games, and online meetings. The VR application contained six scenes (forest, beach, ocean, outer space, city, and indoors) that automatically switched every 30 seconds. Grandparents and grandchildren wore VR headsets to experience these virtual environments together and asked any questions about VR during this part.
 
 \subsubsection{Part Two: VR Experience}
\label{sec:vr-experience}
 \rv{Accompanied by experimenters}, the grandparent and grandchild were guided to two separate rooms with sufficient sound insulation to start the VR experience. During the VR experience, grandparents sit on a chair, and grandchildren stand or sit according to their preference. After completing the tutorials and feeling confident with using the application, participants started by creating an avatar for themselves. Grandchildren customized avatars independently \rv{and can ask for experimenter's help if they need}. Considering their limited experience with VR, the grandparents created their avatars with the help of the experimenter, they picked their favorite appearance (e.g., face shape, hair color, hairstyle, clothes) on pieces of paper, and the experimenter helped them create the avatar in Rec Room.
 
 After creating their avatars, grandparent and grandchild entered a private lounge in VR to communicate and interact with each other \rv{with the mediation by experimenters}. We gave participants two triggers to start their conversation. Still, they can proceed with their own topics as desired: 1) talk to each other about some events that happened recently, (2) interact with surroundings (e.g., playing table tennis or darts, taking physical interaction with each other).

 \subsubsection{Part Three: Semi-structured Interview}
 
 \rv{This part sought to understand participants' communication experiences and preferences of avatar appearance and interpersonal communication in VR. We also asked them about their perceived barriers to using VR as a communication tool in the future. Example interview questions related to this study included: \textit{what avatar appearance did you create? why?}, \textit{how do you feel when communicating with your grandparent/grandchild with these avatars?}, \textit{how you interact with your grandparent/grandchild, why?} and \textit{do you think it is practical to use VR as your future communication tool?} Participants could express their VR communication experiences by comparing their experiences with face-to-face or videoconferencing tools.}
 
 \st{This part sought to understand participants' experiences of communication in VR by comparing them with their experiences of F2F and VC. Participants first took 10 minutes to freely communicate with each other via VC (in two separate rooms) and F2F to evoke their previous communication experiences. Participants then reported their communication experiences in VR, F2F, and VC from five aspects: ease of communication, willingness to communicate, physical intimacy, emotional closeness, and self-report emotions. The ease or willingness to communicate is to evaluate participants' communication quality, and the physical intimacy or emotional closeness is to assess their perceived relational closeness with each other. Grandparents and grandchildren rated their communication experiences on the first four aspects on a 7-point Likert scale on VR, F2F, and VC, and scored their emotions toward these communication approaches on a radar chart.}
 \st{After completing the questionnaire, we interviewed the reasons for their rating on each question by asking: 1) What factors influenced your experience in this communication mode? 2) What are the advantages/disadvantages of VR compared to F2F and VR in this aspect? We also interviewed their motivations of avatar creation to understand their preferences of the avatar appearance.}
 
 \subsubsection{Part Four: Participatory Design}
 
 We conducted a participatory design session to understand their visions of shared activities in VR.
 Firstly, participants were guided to brainstorm shared activities that they would enjoy in VR. Next, the experimenter merged these activities with the eight preset activities generated before the study based on the commonality to form an activities list (It should be noted that all the activities participants proposed were contained in our preset activities). Participants negotiated and sorted activities from the most favorite to the least favorite one, and gave their reasons for their sorting.
 
 Secondly, grandparents and grandchildren chose their favorite shared activity to design further details. Participants started by sketching on paper separately since designing together may cause biases and compromises~\cite{liaqat2021participatory}. We prepared a few questions that probed them to consider detailed characteristics of their design, such as what setting they would like this activity to take place in, and what kind of interaction they wanted. After completing their designs, participants introduced them to each other and then discussed whether the designs were suitable and appealing for the other person. Participants were allowed to modify their designs during the discussion.

\subsection{Data Analysis}


Our data included video recordings of VR experience, interviews, co-design, and the sketch participants drew. The video recordings were first transcribed into a text script. Two authors read through the text script of three randomly selected families several times to understand participants' communication experiences and preferences in VR. Then, they independently coded the script using an open-coding approach~\cite{corbin2014basics}. We combined deductive and inductive coding techniques to form the codebook. First, we have established the four main themes: avatar appearance, interpersonal interaction, shared activities, and the barriers to using VR. We inductively constructed the sub-themes and the specific contents within each main theme by assigning keywords to the participants' feedback or answers. We grouped the repeating keywords at a higher level and formed a diagram with multiple levels. For instance, when participants introduced what motivated them to customize the specific avatar appearances, we would label that part as the main theme ``avatar appearance''. Then, we identified the sub-theme ``avatars reflecting the ideal versions of themselves'' when the words ``younger,'' ``healthier,'' and ``become more confident'' repeatedly appeared in the responses. The inter-rater reliability (IRR) was calculated to be 89\%, which shows a high level of agreement. The two coders regularly discussed the codes and resolved disagreements to create a consolidated codebook. Further meetings were scheduled with all co-authors to reach agreements based on the initial coding result.
 Finally, we generated our tags surrounding the three themes in the research questions. They are \textit{avatar appearance}, \textit{interpersonal interaction}, \textit{shared activities}, and \textit{the perceived barriers of VR as communication media}.

 \sout{Study sessions were recorded and transcribed. We analyzed the qualitative data (including participants' answers to interview questions, verbal descriptions of their sketches, and experimenters' observational notes) using the thematic analysis method~\cite{braun2006using, braun2012thematic}. Two authors first thoroughly read the transcripts of three randomly selected families and then generated their initial codebook using an inductive approach \cite{thomas2006general}. Subsequently, we iteratively discussed these initial codes with the project team in our weekly research meetings and consolidated them into potential thematic areas, then refined these themes through thematic mapping. Finally, we used the refined themes to finish coding the remaining transcripts.}

 \sout{For the quantitative data, we analyzed the participants' ratings of the questionnaire to show their communication experiences in VR, F2F and VC. We plotted the scores of ease to communication, willingness to communicate, physical intimacy, and emotional closeness on the boxplots and drew the average score of their self-report emotion on a radar chart. We ran the Wilcoxon Signed-Rank tests to compare participants' experience of each aspect in VR with F2F and VC.  We presented the quantitative and qualitative results to answer our research questions accordingly in the following section.}

 \begin{figure*}[]
\includegraphics[width=0.9\linewidth]{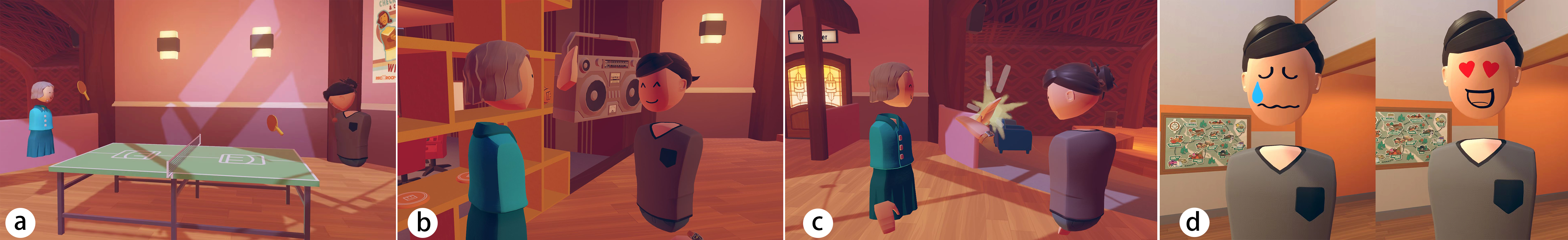}
\vspace{-0.2cm}
\caption{Screenshots of participants' interactions in VR: a) Playing ping pong together, b) Chatting about the radio,c) High-fiving each other, d) Showing emotions using exaggerated expressions.}
\Description{Screenshots of participant interactions in Rec Room: a) Playing ping pong together, b) Chatting about the radio,c) High-fiving each other, d) Showing emotions using exaggerated expressions.}
\label{fig:recroom}
\end{figure*}

\section{Findings}

\rv{We present our key findings on the four themes revealed by our data analysis: 1) grandparents' and grandchildren's preferences for their avatar appearances, 2) their interpersonal interaction in VR, 3) their perceived benefits of participating in shared activities and their envisioned new shared activities, and 4) their perceived barriers to adopting VR in inter-generational communication. }

\st{\textbf{Ratings of communication experiences in VR vs. Other Communication Modes}}

\st{ \textbf{VR vs. F2F}: The results showed that participants were more willing to communicate with each other ($Z= 2.167, p<.05$) and perceived more physical intimacy ($Z= 2.031, p<.05$) in VR than F2F. Compared with F2F, VR was perceived to have less ease of communication ($Z= -2.842, p<.05$) and lower emotional closeness ($Z= -2.224, p<.05$). Participants felt more excited in VR ($Z= 3.503, p<.05$), and grandparents showed more nervousness in VR ($Z= 2.065, p<.05$). }

\st{\textbf{VR vs. VC}: Participants felt more willing to communicate with each other ($Z= 3.171, p<.05$) and perceived more physical intimacy ($Z= 3.822, p<.05$) in VR than VC. There were no significant differences between VR and VC in ease of communication ($Z= 1.352, p=.0740$) and emotional closeness ($Z= 0.932, p=.351$). For participants' reported emotions, they felt more excited ($Z= 3.791, p<.05$) and cheerful ($Z= 2.296, p<.05$) in VR than in VC. The grandchildren felt more relaxed ($Z= 3.141, p<.05$) and less nervous ($Z= -2.146, p<.05$) in VR than VC, while grandparents were more nervous in VR than VC ($Z= 1.794, p<.05$). }

\st{\textbf{Pros and Cons of remote VR communication}}

 \sout{\textbf{P1: Avatars can be customized to reduce age-related power imbalance or convey physical well-being.}}
 
 \sout{Participants believed one contributing advantage of VR over F2F and VC is that VR allows people to customize their own avatars to an abstract one or a realistic one based on their communication needs. The abstract avatars were beneficial for \textit{reducing age-related power imbalance} between grandparents and grandchildren while the realistic avatars help them \textit{get a better sense of the well-being of each other} when living apart.}
 
 \sout{In our study, participants customized their own avatars in Rec Room, which offered abstract representations that did not closely resemble participants' physical features \cite{recroomWebsite}, as shown in Figure \ref{fig:recroom}. We found that abstract avatars could \textit{reduce age-related power imbalance} between grandparents and grandchildren since differences in height and detailed features such as wrinkles were not shown. Grandchildren thought that the avatars of their grandparents looked young and energetic, so they felt that they were no age differences or generation gaps between them (F2, GD:\textit{``My grandpa[avatar] inside was so cute! I really wanted to touch him''}). } \sout{Therefore, grandchildren felt more equal and closer to their grandparents in VR. They were observed to voluntarily touch their grandparents frequently, and even pat their grandparents on the head, which is considered inappropriate in families with conservative cultures~\cite{kennedy1992shared, kemp2005dimensions}. Participants stated that it was totally different when interacting F2F --- grandparents often touched their younger relatives to express their intimacy, while the grandchildren would not touch their grandparents casually (F6, GD: \textit{``We need to have a sense of propriety and respect for the elderly''}). When asked what they would do if the avatars had the same face as their grandparents, the grandchildren thought that they would not touch them so frequently or casually.}
 
 \sout{In addition to the benefits of abstract avatars, we uncovered potential advantages for realistic avatars (those that closely resembled the physical appearances of the user) in remote inter-generational communication in the interviews. Participants felt that realistic avatars would help them \textit{get a better sense of the well-being of each other} when living apart. So that they would check up on their physical health and emotional well-being (F8, GD \textit{``We would always care about each other's health when we contacted''}), and satisfy their longingness for each other, especially after being separated for a long time. }

\subsection{Grandparents' and Grandchildren's Preferences for Their Avatar Appearances}
\rv{Grandparents and grandchildren wanted themselves to be represented by three main types of avatars while communicating in VR: 1) \textit{avatars reflecting their current physical appearances}; 2) \textit{avatars reflecting the ideal versions of themselves}; 3) \textit{avatars with no obvious reference to themselves}.}
 

 \subsubsection{ \rv{Avatars Reflecting Their Current Physical Appearances}}
 Grandparents (N=5) and grandchildren (N=7) offered four advantages of using the type of avatars. First, they felt this type of avatar allowed them to convey their physical and spiritual well-being to each other. Participants mentioned that when living apart, they expected to get a better sense of the well-being of each other through their physical appearance, as one participant (F8, GD) elaborated: \textit{``We always care about each other's health when we communicate.''} Second, participants felt more relaxed as they did not need to worry about disguising themselves or dressing themselves up when using this type of avatar, as one participant (F11, GS) explained, \textit{``it is natural to meet my grandmother with my real appearance''}. It is worth noting that many grandchildren participants emphasized that they would choose different styles of avatars when communicating with their young friends or strangers, such as more extravagant or exaggerated avatars (e.g., more exaggerated hairstyles or dress up, healthier or stronger body) to make themselves stand out. 
 \rv{Third, participants felt that they could start conversations and interactions more easily with this type of avatar due to the familiar appearances of the avatars. Lastly, they also commented that compared to the other two types of avatars, this type of avatar could better offer a sense of companionship to relieve their nostalgia when living apart. }

 \rv{In addition to the above four advantages of this type of avatar, participants also mentioned differences between using this type of avatar in VR and using videoconferencing tools even though both approaches allowed them to see each other's current physical appearances. On the one hand, VR allowed participants to communicate in an embodied manner through avatars. Such embodied avatars in virtual 3D environments allowed them to enjoy a sense of companionship and expressed themselves via various non-verbal cues, such as changing their personal distance or their facial orientation and pointing to direct attention. Such experiences were more like physical world interactions but were largely unavailable when communicating via videoconferencing tools. One participant (F12, GS) elaborated: \textit{``In video calls, we are limited to what the camera frame captures. We cannot see or feel the entire environment where the other one is talking in.''}  
 On the other hand, 
 videoconferencing tools allowed them to see each other's facial expressions but current VR apps could not, which made it difficult to perceive each other's emotions, as one participant (F2, GD) explained: \textit{``It [VR] is more like a phone call in the sense that I can only feel my grandpa's emotion through his voice''}. Similarly, while eye contact is important for showing respect and attention, it was mostly unavailable with current VR devices, as one participant (F5, GS) explained: \textit{``it [eye contact] is a way to show my respect and attention in this conversation.''}} 

 \subsubsection{\rv{Avatars Reflecting the Ideal Versions of Themselves}}
 
 When considering creating avatars reflecting the ideal versions of themselves, grandparents (N=6) and grandchildren (N=3) had different preferences for their ``ideal versions''.  \rv{Grandparents tended to want their avatars to \textit{look like their younger selves}. 
 Grandparent participants expressed three main reasons why they would want their avatars to look like their younger selves: \textit{to reminisce their past experiences (N=5)}, \textit{to show their youth's appearances to their grandchildren (N=4)} and \textit{ to gain confidence (N=3)}.} One participant (F6, GP) elaborated: \textit{``I chose to wear military uniform [in VR] because I always wore it when I was young...I was strong at that time, but you [granddaughter] have never seen it...''}. These grandparent participants depicted their avatars with words like \textit{taller}, \textit{stronger}, \textit{black hair}, and \textit{without a wrinkle on skin}.
 When represented by avatars with their younger appearances, grandparent participants felt more confident and energetic than in reality and were more willing to engage in activity and conversation with their grandchildren. One participant (F7, GP) elaborated: \textit{``I like the feeling of being young, and wants to play with her [grandchild] like young people.''} 

 From the grandchildren's perspectives, they felt pleased to see their younger grandparents in VR and thought these avatars helped them to blur their generation gaps. Communicating with grandparents through avatars of their younger appearances in VR, grandchildren also felt more equal to grandparents than in reality. one participant (F10, GS) elaborated: \textit{``In VR, he [grandfather] was like my friend, rather than my elder family.''}. Such a blur of generations encouraged them to contribute to their conversations and allowed them to have more fun in their shared activities.
 \rv{Additionally, grandchildren also felt that their grandparents' avatars with their youth appearances triggered their curiosity to want to learn more about their grandparents' old times, which sparked conversations about the past and reminiscence.} One participant (F9, GD) depicted this experience: \textit{``She [grandmother' avatar] has two big black braided hairs in VR, but actually she has short hair in reality. I wondered why she chose this hairstyle. And she told me that she used to have long hair when she was young.''} 

 \rv{On the other hand, grandchildren tended to want their avatars of ideal selves to have \textit{more extravagant and good-looking appearance} than what they would normally look like in real life by adjusting their physical traits or dressing up better, such as dressing up like a wizard or having long hair that they dreamed for. }
One participant (F9, GD) elaborated: \textit{``I made my avatar thinner and dressed in a style that is impossible to try in daily life''}. Being represented by the avatars of their ideal selves, they also perceived more confidence when interacting with their grandparents in VR. Grandparents were also generally receptive to this type of avatar and even felt fun to meet with their dressed-up grandchildren, as another participant (F2, GF) explained: \textit{``she's getting cooler, isn't she?''}

 \subsubsection{\rv{Avatars with No Obvious Reference to Themselves.}}
 To a lesser extent, participants also represented themselves with the type of avatars that had no obvious reference to themselves. They felt crafting avatars of this type as a \textit{dress up game}, that would allow them to have fun together, as one participant (F1, GS) explained: \textit{``I chose this black beard and a strong body to look like a hunter because I thought it was fun to use it to communicate with my grandparent.''}. \rv{One drawback participants mentioned was that it might take more time for them to get used to this type of avatar and start conversations, as one grandparent (F8, GP) elaborated: \textit{``I felt she's a stranger...or a character in a video game...It's weird talking to her.''}}

\subsection{\rv{Interpersonal Interaction in VR}}

 \rv{Interpersonal interactions via embodied avatars showed two advantages of VR in facilitating communication between grandparents and grandchildren: 1) \textit{intimate physical contact and inappropriate interactions in real lives become more acceptable};} 2) \textit{richer sensory feedback may help to provide a stronger feeling of togetherness and intimacy}. 
 
 
 \subsubsection{Intimate physical contact and inappropriate interactions in everyday lives become more acceptable in VR} 
 \rv{Participants (N=7) made physical contact, including holding hands and hugging, frequently when interacting with each other in VR via embodied avatars. While interpersonal interactions with intimate physical contact may be common in many western societies, they are often considered too sentimental and rarely happen in eastern societies, where our participants grew up and lived. One possible reason for this change might be that VR gamified physical touch and hugs and made them feel more casual and natural. One grandchildren participant (F1, GS) elaborated: \textit{``I rarely perform intimate interactions with my grandma in real life because it feels so strange and embarrassing. Surprisingly, in VR, touching and hugging her became natural, and I felt I was closer to my grandma.''} 
 

In addition, participants, especially grandchildren, performed interpersonal interactions, such as patting their grandparents' heads and poking their grandparents' cheeks, which would be considered ``inappropriate'' or ``impolite'' in the society where our participants lived. Participant (F6, GD) elaborated \textit{``We are educated to have a sense of propriety and respect for the elderly.''} Interestingly, these ``inappropriate'' interactions became acceptable and even fun to perform in VR. The younger and cartoon avatars in which they embodied blurred the generation gap between grandparents and grandchildren and allowed grandchildren to gain fun and show intimacy. Moreover, grandparents also showed high tolerance toward these interactions, as one participant elaborated:\textit{``I like the way how my grandson interacts with me, and it makes me feel like we are peers.''} 
Furthermore, doing such ``inappropriate'' interactions in VR could potentially help both generations get used to them in real lives, as one participant (F12, GS) argues:\textit{``We were not used to expressing love by hugging each other. However, it felt easy and natural to hug her in VR...I might get used to hugging her.''}}
 

 \subsubsection{Richer sensory feedback may help to provide a stronger feeling of togetherness and intimacy.} In addition to being embodied in avatars and performing interpersonal interactions, participants (N=17) felt that sensory feedback beyond vision and audio could help enhance their feeling of togetherness and intimacy. For example, although current VR handles can only provide simple vibration as haptic feedback, many participants already felt it enhanced their feeling of togetherness, as one participant (F5, GF) explained: \textit{``Although I knew it [vibration] was just a virtual perception instead of a real touch [on my granddaughter], it still made me feel like that she is right beside me.''} 
In addition, participants expected to have richer sensory feedback to provide them a feeling of meeting ``in person'' and alleviate their feelings of nostalgia. For example, one participant (F9, GM) elaborated: \textit{``We haven't seen each other for three years because she went abroad. I really hoped to be able to sit and chat with her face-to-face and pat her on the back. I hope that VR can provide the real feeling of touching, which may be a gift for the elderly like us who don't see our family members often''}. 
Regarding richer sensory feedback, participants hoped to experience the tactility of textures and materials in VR so that they could get the feeling of touching each other's hair and skin or being touched or hugged. 

\subsection{Perceived Benefits of Participating in Shared Activities and Envisioned New Shared Activities}

\rv{We first present two potential benefits of engaging in shared activities in VR and then elaborate on three kinds of shared activities that participants envisioned doing together in VR.}

 \subsubsection{Benefits of Engaging in Shared Activities in VR}
 Engaging in shared activities in VR had two potential benefits for fostering positive experiences between grandparents and grandchildren. First, engaging in shared activities in VR stimulated more topics for them to keep their conversations flowing. For example, one participant (F3, GM) elaborated: \textit{``There's just a big talking head in the video call. Sometimes we stare at each other and don't know what to talk about...However, playing ping pong together or watching him draw [in VR] triggered more topics to chat about. For example, we chatted about how he learned table tennis as a kid...''}

 Moreover, engaging in shared activities helped grandparents and grandchildren avoid awkward silent moments, which are common during videoconferencing as participants indicated. They often do not know what to do when silence inevitably comes in videoconferencing and tend to end their conversations earlier than they would otherwise want. One participant (F6, GF) explained:\textit{``Video call is like a Q\&A session. One person asks a question, and the other answers it. If we can't think of a question for a while, we just hang up...''}). In contrast, in VR, engaging in shared activities allowed the grandparents and grandchildren to fill in the inevitable silence by shifting their attention toward the activity that they could do together, as one participant (F3, GS) elaborated: \textit{``it (VR) gave you something to key into... We didn't feel embarrassed when we have nothing to say, we just continued playing ping pang.''} While engaging in the shared activity, they could further identify new topics to keep their conversations flowing.
 

\subsubsection{Participants' Envisioned Shared Activities for Themselves to Engage In}
\begin{figure*}[tbh]
\includegraphics[width=0.9\linewidth]{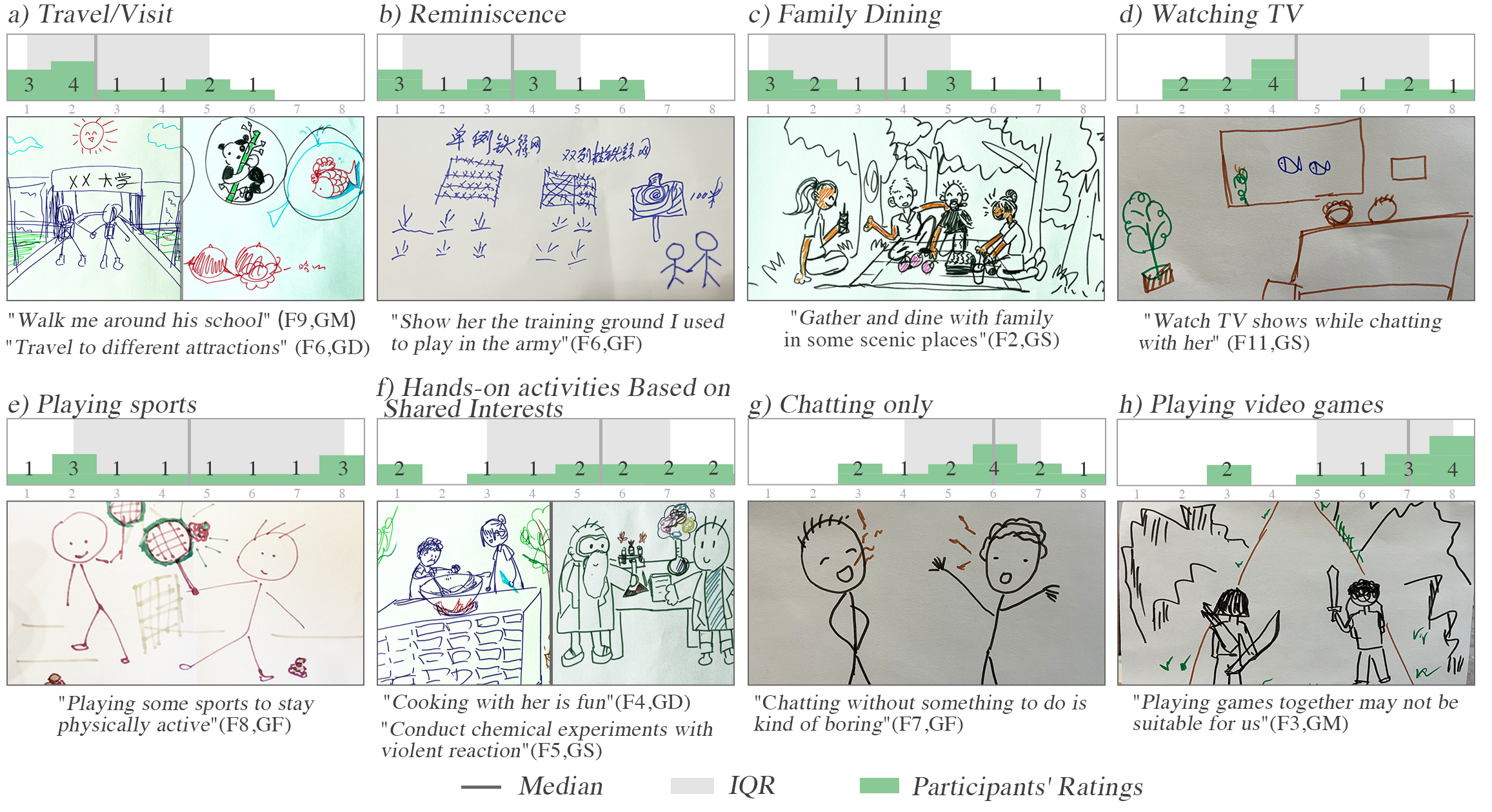}
\vspace{-0.2cm}
\caption{Participants' rankings of the eight shared activities in VR with examples sketches from Part 4 of the study for each activity (1: most favorite, 8: least favorite) }
\Description{Participants' rankings of the eight shared activities in VR with example sketches from Part 4 of the study for each activity. a) Travel/visit, b) Reminiscence, c) Family meeting, d) Watching TV, e) Playing sports, f) Hands-on activities based on shared interests, g) Chatting only, h) Playing video games}
\label{fig:activities}
\end{figure*}

Figure~\ref{fig:activities} shows participants' eight envisioned shared activities with example sketches, which are ranked based on grandparents' and grandchildren's preference ratings shown above the sketches. We further categorized these activities into three groups based on their similarities: \textit{exploring difficult-to-visit locations}, \textit{reminiscing together}, and \textit{re-experiencing daily activities}. 
 
 \textbf{Exploring difficult-to-visit locations.}
 First, our participants showed their willingness to experience difficult-to-visit locations in the physical world together. Frequently mentioned locations include personally meaningful places and tourist attractions, as shown in Figure \ref{fig:activities}a. For the personally meaningful places, participants argued that VR could help grandparents, especially those who have difficulty walking, visit faraway residences and schools of their grandchildren. With such support, grandchildren could show their living surroundings and share their life stories with their grandparents more vividly. One participant (F8, GM) elaborated: \textit{``I have never gone to her college due to the pandemic's situation. I hoped that she can walk with me around her college.''}. Exploring such difficult-to-visit locations may enhance grandparents' understanding of grandchildren's lives and offer more common topics in future conversations. 
 For the tourist attractions, participants felt that VR could allow them to enjoy the beautiful scenery and share their perceptions with each other without worrying about various concerns of traveling in real life, such as time, distance, language barriers, and personal health. 
 One participant (F5, GF) explained: \textit{``we can just enjoy the beautiful sightseeing and do not need to spend a long time commuting.''} Another 86-year-old participant, who had difficulty traveling due to his poor physical health, commented: \textit{`I haven't traveled far in about ten years because of my physical issues. I am so bored of staying home... With VR, I can experience the world `in-person', not through TV''}.



\textbf{Reminiscence together.}
 The second category of shared activities was reminiscence, as illustrated in Figure \ref{fig:activities}b.
 Grandparent participants wanted to share their personal experiences and stories in the past with their grandchildren. One common aim was to pass on their values and ideologies to the younger generation, as one participant (F1, GF) elaborated: \textit{``My past experiences of dealing with difficult times may help her gain a sense of responsibility and encourage her to be a hard-working person.''} On the other hand, grandchildren participants felt that family stories were enjoyable as they could provide complementary perspectives to understanding what they learned from books, as one participant (F2, GD) explained: \textit{``It makes me feel that the events in history books were closely related to the development of our family.''}
 
 Grandparent participants pointed out that one common approach to sharing their stories with their grandchildren was just talking or talking with the assistance of some items, such as old photos. with this approach, grandchildren often have to use their imagination to reconstruct story elements that they may not have experienced, as one participant (F4, GD) elaborated: \textit{``I have never seen the scenes that grandma described in the stories. It is hard to imagine.''} Participants felt that VR might be able to provide them with 3D immersive environments of old times together. Such environments not only could offer more appealing and detailed visuals for grandparents to tell stories but also allow grandchildren listeners to physically engage in the story by interacting with surroundings (F7, GD: \textit{``My grandpa always tells me his hunting stories but I have no idea of that experience. With VR, I can hunt in the mountains with my grandpa by pulling a bow and shooting a wild boar.''})

 \textbf{Re-experiencing familiar daily activities.}
 The last category of shared activities that participants wanted to engage in was re-experiencing familiar daily activities that they used to do together in reality (e.g., family dining, watching TV). Through familiar daily activities, participants felt they could better enjoy the company of each other and get used to new VR operations more easily. 
 Furthermore, participants imagined that VR could potentially revive their experiences of these familiar daily activities with rich and spectacular visual effects, as one participant (F5, GF) mentioned: \textit{``we can do some dangerous and violent chemical experiments that cannot be done in real life''}.
  

 Participants envisioned four common types of familiar daily activities (see Figure \ref{fig:activities} c, d, e, f, g):  \textit{family dining}, \textit{watching TV}, \textit{playing sports} or \textit{doing hands-on activities}, and \textit{simply chatting}. The most common familiar daily activity was dining together as they did in real life, which  is a familiar social activity for family members to meet and share experiences. 
 Another familiar daily activity was watching TV together in VR, which is one of the most common in-person activities between grandparents and grandchildren in real life~\cite{kennedy1992shared}. Participants felt it did not require complex operations in VR. However, they also felt it less attractive than traveling to other places. For example, one participant (F11, GM) elaborated: \textit{``I think it is easy [to watch TV in VR] because we don't need to learn to operate much. But I prefer to use VR to do something that can't be done in reality''}. The third familiar daily activity was playing sports together. Many participants played table tennis together in the private lounge in the VR experience session and felt operations for doing sports in VR were easy and intuitive, such as swinging the racket to hit the ball. 
 Furthermore, some participants hoped to engage in hands-on activities that they often undertake in real life based on their mutual interests and hobbies. Examples of hands-on activities include cooking, playing cards, making crafts, and conducting science experiments. However, participants felt these activities are intricate and often have high operational difficulty with hand controllers, especially for the elderly. This resulted in lower rankings since grandparents felt it could be exhausting to do such activities in VR. The last family daily activity to engage in was simply chatting. Though it was less cognitively and physically demanding, participants felt it might be boring. If they had to chat only, they hoped to have some prompts to motivate them to strike up conversations.

 In addition to these three categories of shared activities, participants mentioned playing video games in VR (e.g., shooters or action-adventure). Although more than half of the grandchildren mentioned that playing video games would be on the top of their personal favorite list, they thought it was not a suitable activity to do with their grandparents, as one participant (F10, GF) explained: \textit{``We have no habit of playing games together. I don't think it is necessary to spend time learning it.''}
 
\subsection{\rv{Perceived Barriers to Adopting VR for Grandparents and Grandchildren Communication}}

 \rv{While participants were generally positive about VR's possibility to facilitate intergenerational commendation between grandparents and grandchildren, they also shared three perceived barriers to VR for intergenerational communication. One prominent barrier was the overhead of initiating communication in VR (N=16) especially compared to other media, as one participant (F11, GS) explained: \textit{``We can chat via phone call by simply turning on the phone and dialing. But to communicate in VR, we have to first find a physical space, wear the headset, and set up the application before we can meet...''}
 Despite the overhead of using VR, it was still perceived to be more appealing than video calls because it could provide a sense of togetherness and relieve nostalgia, especially for grandparents and grandchildren who have separated for a long time (e.g., grandchildren who went abroad for school). One participant (F9, GM) elaborated: \textit{``[In VR] I felt like I was sitting and chatting with her face to face. It's amazing... I hope to use it to contact my family members living far away from me''}. }
 
 The second barrier was the perceived difficulty of the operations of VR controllers, especially for grandparent participants (N=8).
 Many grandparent participants found some operations (e.g., pushing the ``A'' button to navigate) non-intuitive and difficult to learn and thus had to spend more energy and attention learning such operations, which made them nervous and distracted them from communicating with their grandchildren. However, grandparent participants were receptive to the operations that simulate their familiar actions in real life, such as swinging the controller to hit the ball and clapping the other's hands. During the VR experiences, grandparents (N=10) asked their grandchildren for help via voice in VR, and their grandchildren often patiently taught the grandparents how to operate the hand-controller to operate.
 
 \rv{The last barrier was related to safety concerns (N=5). Many participants worried that they might bump into surrounding objects and fall because VR headsets covered their faces and blocked their sight. They pointed out that this might be particularly dangerous to older adults who live alone, as one participant (F12, GS) explained: \textit{``I'm worried about letting her use VR at home alone because I can't help her if she falls.''} The VR headset we used (Quest2) allowed users to define the ``play area'', which shows a virtual fence and reveals the real surroundings to the VR user if she steps out of this area. However, participants still felt it insufficient and hoped to check their surroundings from time to time as such a ``play area'' cannot reveal moving objects, such as smart vacuum cleaners that move around. Consequently, participants felt that they would only be comfortable using VR if safety concerns are out of the way.}

 \sout{\textbf{Demographic Factors that Affect VR Experiences between Grandparents and Grandchildren}}

 \sout{According to our study, several factors affect participants' perceptions of VR-based remote communication, such as in-person meeting frequency, the closeness between grandparents and grandchildren, the gender of the grandparent, and the health condition of the grandparent.}

 \sout{We found that the \textit{in-person meeting frequency} affected their willingness to use VR for remote communication: the fewer in-person meeting frequency and the longer of separation, the more they want to communicate using VR. }
 Because of its simplicity and familiarity, they tended to use phone calls to contact each other when the separation first arises. However, for those who have separated for a long time (e.g., grandchildren who went abroad for school), VR became more appealing because the sense of co-presence and togetherness brought by VR could relieve their nostalgia. 
 
 \sout{\textit{The closeness between grandparents and grandchildren.} We measured the closeness between grandparents and grandchildren using the frequency of remote communication and the length of time they lived together~\cite{geurts2012grandparent, viguer2010grandparent}. Although all participants were cheerful and excited when communicating in VR, the grandparents and grandchildren who had closer connections exhibited a higher acceptance of VR as their remote communication tool in the future. The grandparents and grandchildren with weaker bonds thought communicating in VR with each other was more burdensome.}
 
 \sout{Different \textit{gender of grandparents} may play different roles in the grandparent-grandchild relationship \cite{kennedy1992shared, sheehan2008grandparents}, which affect their favorite VR activities. Participants who were grandfathers tended to share stories from their youth through VR activities, hoping that their grandchildren would learn some knowledge and lessons. Participants who were grandmothers preferred to have a family meeting or travel in VR, enjoying the happiness of chatting together with family members.}

 \sout{\textit{Health condition of the grandparent.} Our results showed that elderly participants with poor health were more likely to accept VR. Two of our older participants had difficulty walking, and had not traveled for a long time. They appreciated the immersion brought by VR, which made them feel like they actually visited different places, and enabled them to visit their grandchildren ``physically''. They felt more cheerful and excited in VR than the healthy participants.}

\section{Discussion}

\rv{We investigated how VR could be leveraged and designed to facilitate inter-generational communication between grandparents and grandchildren through a participatory design study in which participants interacted with each other in VR and designed features to show how they would want VR to be designed to facilitate their communication. Our findings show that 1) participants chose different types of avatars to satisfy different communication needs; 2) interpersonal interactions via embodied avatars, especially those that have intimate physical contact or are considered ``inappropriate'' between grandparents and grandchildren in real life, could facilitate inter-generational communication; 3) participants felt that engaging in shared activities allowed them to find more topics to fill in the silence in conversations and further envisioned three types of shared activities they wanted to engage in; 4) participants also perceived three main barriers of adopting VR for inter-generational communication: overhead of setting up VR for communication, perceived difficulty of certain VR operations, and safety concerns.  
In this section, we further discuss the design implications of our findings and related future research agenda for better leveraging VR to facilitate inter-generational communication between grandparents and grandchildren. }

\sout{\textbf{Role of VR for Grandparent-grandchild relationship} }

 \sout{We identified two roles VR played in remote inter-generational communication based on the findings, which are \textit{communication facilitator} and \textit{nostalgia reliever}.}
 
 
 \sout{Results show that VR acted as a \textit{communication facilitator} in three ways: allowing  \textit{the customization of avatars}, providing\textit{shared activities}, and offering \textit{a wide variety of non-verbal language}.  First, VR enabled participants to use abstract avatars, which encouraged grandchildren to talk and physically interact with their grandparents more casually and frequently. Thus, VR made grandchildren feel more equal with their grandparents by reducing age-related power imbalance.
 Second, through engaging in shared activities provided by VR, grandparents and grandchildren gained more fun and exchanged their thoughts and emotions more easily than only speaking. Additionally, shared activities helped fill in the inevitable silence during conversation and provided more opportunities to strike more conversations.
 Third, VR enriched communication by offering a wide variety of non-verbal language, enabling participants to express themselves more effectively.}
 
 \sout{VR also acted as the \textit{nostalgia reliever} for remote grandparents and grandchildren to alleviate their feelings of nostalgia, especially for those who have been separated for a long time. \textit{The simulated physical interactions} enabled remote users to touch each other, while \textit{co-presence} provided a sense of intimacy despite the long distance between them in real life. Additionally, the emerging techniques of real-time 3D reconstruction \cite{orts2016holoportation, izadi2011kinectfusion, newcombe2015dynamicfusion} may allow grandparents and grandchildren to use realistic avatars. Therefore, remote grandparents and grandchildren could interact in the same virtual room with real faces to show their well-being and physical health, which participants believed can satisfy their longingness for the other.}
 
 \sout{Overall, combined with previous research indicating that using VR for communication can improve the quality of life for older adults and their remote families~\cite{afifi2022using}, our results show that VR is a viable communication media for remote grandparents and grandchildren by supporting the need of communication and relieve nostalgia. }

 \sout{\textbf{The Uniqueness of VR Communication for Grandparent-grandchild Relationships}}
 
 \sout{As discussed in Section 2, no prior work investigated remote communication in VR for grandparent-grandchild relationships. Our work uncovered the uniqueness of VR communication in a grandparent-grandchild relationship compared to other relationships (e.g., friends, couples, or strangers) by revealing several differences between these contexts.}

 \sout{Although the pros and cons of VR in communication may be applicable to a wide range of people, they contained nuanced differences for specific types of relationships. }
 \sout{For \textit{the customization of the avatar}, abstract avatars in different relationships made participants feel relaxed and cheerful during conversation~\cite{baker2021avatar, Li2019, maloney2020virtual}, while it played a specific role in reducing the age-related power imbalance for grandparent-grandchildren relationships. In contrast, when communicating with strangers, people enjoyed the sense of anonymity of the avatar and felt encouraged to have more social interactions with others~\cite{baker2021avatar, kang2013impact}.  For \textit{the feeling of co-presence and simulated physical interactions}, grandparents and grandchildren regarded it as a way to alleviate their feelings of nostalgia and longingness, while friends or strangers saw it as a novel experience and interesting interaction~\cite{kim2014improving, maloney2020virtual}. Pairs with different relationships (e.g., friends, strangers, grandparent-grandchild) gained entertainment from \textit{the wide range of activities and novel environments} offered by VR~\cite{freeman2022working, freeman2021hugging, maloney2020virtual}. However, grandparents and grandchildren valued the ability of activities and environment to encourage conversation, while friends pay more attention to the playfulness of the activities and environment\cite{freeman2021hugging, maloney2020virtual}. }
 \sout{The \textit{lack of facial expressions and emotional expressivity} is a general disadvantage that negatively impacts people's communication experiences in different relationships~\cite{Li2019, baker2021avatar, smith2018communication}. Yet, the tendency of grandchildren to closely monitor the emotions of their elders is unique to grandparent-grandchildren relationships, because of the pressure and responsibility they feel~\cite{kemp2005dimensions}. Therefore, lacking emotional expressivity may introduce more barriers to communication between grandparents and grandchildren than others. Another con for engaging in VR activities is the \textit{operation difficulty}, which may be amplified in grandparent-grandchild relationships because it distracted grandparents' attention from communication and made them feel nervous and anxious. However, it is no longer a significant shortcoming when interacting with friends --- they enjoy the challenge and fun brought by the various operations~\cite{tan2011dynamic}.}
 
 \sout{For the \textit{avatar preferences}, our results show that grandparents and grandchildren felt more relaxed and spent less time on avatar creation in inter-generation communication because they felt at ease when staying with their family. In contrast, they would craft their avatar more carefully when communicating with friends or strangers ~\cite{kolesnichenko2019understanding, Tanenbaum2020}. Additionally, our participants hoped to use a realistic avatar to show their current status and physical well-being with their grandparents/grandchildren but exhibited less interest in showing their real faces to their friends or strangers.  }
 
 \sout{For the \textit{preferences of shared activities in VR}, grandparents and grandchildren prioritized communication of daily experiences rather than entertainment. They wanted to reconstruct daily activities with novel approaches in VR, which could not only evoke a feeling of togetherness but also facilitate conversation. Previous work also revealed similar needs in long-distance couples~\cite{zamanifard2019togetherness}. However, current commercial social VR applications only provide some common scenarios and activities for general users to have fun~\cite{bigscreenWeb, recroomWebsite}. Future researchers should provide more customized scenarios and activities to meet the communication needs among different relationships.}
 
 \sout{In sum, we discussed significant differences in VR communication between grandparent-grandchild relationships and other types of relationships. Future designs for VR-based inter-generational communication systems should consider the uniqueness of this relationship and its corresponding needs.}

\subsection{Choosing Suitable Avatars for Different Communication Needs}
 
 \rv{Avatars play an essential role in VR by shaping people's communication experiences~\cite{garau2001impact, inkpen2011me}. Our work shows that grandparents and grandchildren customized three kinds of avatars to represent themselves: 1) avatars reflecting the ideal versions of themselves; 2) avatars reflecting their current physical appearances; 3) avatars with no obvious reference to themselves. We found that these avatars brought different communication experiences for grandparents and grandchildren, which inspires future VR applications to provide appropriate avatar appearances to fulfill the unique needs of different communication scenarios. }

 \subsubsection{Design Implication 1: Provide avatars that reflect current physical appearances to check up on each other's well-being and offer a sense of companionship for grandparents and grandchildren who live apart.} 
 People have a strong need and desire to stay connected with families to share their recent life and well-being when they have been separated for a long time~\cite{mynatt2001digital, tee2009exploring, romero2007connecting, neustaedter2006interpersonal}.
 Although prior research explored real-time 3D reconstructions that can relive the details of the remote families' facial expressions, body movements, and gaze on avatars~\cite{orts2016holoportation, li2015facial, schwartz2020eyes}, the uncanny valley effect---where people feel uncomfortable and scared of nearly realistic human appearances~\cite{seyama2007uncanny}---would negatively impact users' communication experiences. In contrast, the abstract avatar may help avoid the uncanny valley. \rv{However, it is still unknown which physical features should be recreated through avatars in VR to fulfill the need to know each other's physical health. Future work should explore essential features that need to be reconstructed on avatars to better show grandparents' and grandchildren's well-being (e.g., figure, complexion, expression) as well as the kind of visual effects needed to show these features on avatars. }


 Embodied avatars in VR can provide a sense of co-presence that make people feel like being together~\cite{smith2018communication, biocca2003toward}. Our study extends prior work by showing that the ``familiar'' appearances of embodied avatars in VR could offer a sense of companionship for grandparents and grandchildren while alleviating their loneliness when living apart. Communicating with the familiar and embodied avatars of each other combines the benefits of traditional video communication media and physical in-person meetings~\cite{Tanenbaum2020, smith2018communication, bailenson2004transformed}. There are many opportunities to leverage the companionship from grandchildren through embodied avatars for older adults suffering social isolation, which negatively affects their health and well-being~\cite{national2020social}. Future work can explore using resemblant appearances on avatars to connect older adults with their remote grandchildren or other family members to improve their life quality.

 \subsubsection{\rv{Design Implication 2: Provide grandparents' avatars with youthful appearances as a stimulus for conversations about the past.}} 
Many participants expressed a positive attitude towards the youthful appearances of their grandparents' avatars because it blurred the generational gaps between them but also triggered grandchildren's curiosity about their grandparents' past. Therefore, youthful avatars can be useful for reminiscence between grandparents and grandchildren by fostering conversations and increasing playfulness. Future VR developers should investigate ways to customize youthful avatars for older adults, for example, by generating avatars based on photos of old people when they were young~\cite{hu2017avatar}.

\subsection{Designing Interpersonal Interaction to Promote a Stronger Sense of Equality and Closeness}

 Our study shows that grandparents and grandchildren found it easy to perform intimate physical contact and interactions that are often considered inappropriate in the real world. Participants felt more equal with each other in VR because of the relaxed atmosphere brought by the gamified environment and embodied representation through avatars. These findings contribute to the following two design implications.

 \subsubsection{\rv{Design Implication 3: Design VR communication platforms for grandparents and grandchildren to communicate and interact more equally.}} \rv{Prior work indicated that although some people regard their grandparents as their friends~\cite{kornhaber1981grandparents}, the admiration and respect for the older generation induce pressure and a feeling of responsibility toward them~\cite{kemp2005dimensions}. Therefore, many grandchildren participants indicated that they seldom speak candidly and interact casually with their grandparents in real life, which leads to less communication and a deeper generation gap between them. In contrast, they felt more equal and easier to perform ``intimate'' or ``inappropriate'' interactions with their grandparents in VR. Additionally, grandparents enjoyed the feeling of being treated as friends in VR. This equal way of getting along may promote communication between the two generations. Therefore, we believe in the potential of using VR as an equal communication platform to encourage them to communicate and interact more than in reality. Future work could explore whether ``equal interaction'' can contribute to closer inter-generational relationships and what kinds of ``inappropriate'' interactions are more acceptable in this relationship to better design a VR communication platform for grandparents and grandchildren.}

 \subsubsection{Design Implication 4: Provide multi-sensory feedback in VR to gain a better sense of togetherness.} As mentioned above, the pressure and responsibility toward grandparents can lead to grandchildren rarely making close physical contact with their grandparents in reality (e.g., hug, kiss or pat), which has negative impacts on building the close inter-generational relationships~\cite{svetlik2005declines}. Our study shows the potential of using VR as a tool for grandparents and grandchildren to experience the physical interactions that they used to feel ``too sentimental'' or ``inappropriate'' to perform in reality.
 Additionally, our work shows that rich sensory feedback was perceived to be helpful to increase the sense of intimacy, which caters to the needs of physical and emotional closeness among remote family~\cite{fuchsberger2021grandparents, zamanifard2019togetherness}. Within the HCI community, researchers have been investigating various sensory feedback, such as haptic feedback~\cite{pamungkas2016electro, ku2003data, burdea1996force, han2020mouille}, smell~\cite{kerruish2019arranging, cheok2018virtual, ramic2010virtual} and sensation~\cite{jiang2021douleur}. 
 Future work should explore what kinds of sensory feedback grandparents and grandchildren would like to perceive in VR and how to simulate such sensory feedback for them. 

\subsection{Providing Rich Shared Activities to Foster Communication}

 Experiencing shared activities helps people exchange their thoughts and emotions naturally while having fun~\cite{viguer2010grandparent, richter2018relations, evjemo2004supporting}. Our participants felt that shared activity in VR acts as a communication facilitator by generating more conversation topics and avoiding awkward silences, which could result in more natural and longer communication.  
 Sufficient communication can help grandparents and grandchildren understand each other better, increase social engagement and reduce social isolation for grandparents \cite{lindley2009desiring, seifert2020digital}.
 In our research, grandparents and grandchildren imagined three types of shared activities that they would like to do together in VR, including exploring difficult-to-visit locations, re-experiencing daily activities, and reminiscing together.

 \rv{Previous work suggested that VR may provide older adults with opportunities to \textit{travel to difficult-to-visit locations} virtually and to enrich their life experiences, especially for those who are in poor physical and cognitive health~\cite{thach2020older, 10.1145/3470743, houwelingen2021virtual, baker2020evaluating}. Our participants suggested that involving grandchildren in such shared activities is appealing for both generations because it allows them to enjoy the fun of traveling together and generate topics to sustain their communication. Our study also found that \textit{the school or home of grandchildren} are desired locations for grandparents to visit to better understand their grandchildren's life. Currently, grandchildren show their living surroundings to their grandparents through video calls~\cite{forghani2014routines, fuchsberger2021grandparents}. VR can augment this sharing experience by providing a 3D reconstruction of grandchildren's entire surroundings~\cite{orts2016holoportation}, which enables grandparents to virtually visit and experience the living environment of their grandchildren.}

 \rv{Our work also found that grandparents and grandchildren would love to \textit{reminiscence} together in VR. 
 Reminiscence was shown to be helpful to improve grandparents' and grandchildren's mood, memory, social engagement, and quality of relationships~\cite{hsieh2003effect, stargatt2022digital, kennedy1992shared, fuchsberger2021grandparents}. Despite these benefits, our grandchildren participants expressed that the lack of common understanding about the objects that their grandparents mentioned, and the lack of interaction while listening, negatively impacted their current reminiscence experiences. They envisioned that VR may potentially enrich the reminiscence process and make it more engaging by recreating the story's setting and allowing them to interact with virtual objects (e.g., hunting with a bow and arrow, or driving a vintage car). Recent research began to show that reminiscence in VR could help older people enrich their life and recollect forgotten memories~\cite{baker2020evaluating, baker2021school, baker2019exploring}. Thus, future work could explore how to offer better reminiscence experiences between grandparents and grandchildren in VR to enrich this collective reminiscence process.}

 \rv{Another key finding of our study was the desire for \textit{re-experiencing familiar daily activities} between grandparents and grandchildren in VR, which could help them better enjoy the company of each other and get used to new VR operations more easily. This finding is in line with prior work that showed that activities familiar to older adults could encourage them to accept VR technology easily~\cite{wang2022construction}. Moreover, our work uncovered three common daily activities that participants would like to engage in: \textit{family dining, watching TV, and playing sports}. Future work could investigate how to reconstruct these activities in VR, and whether these activities could improve the quality of life and relationships for grandparents and grandchildren in the long term.}

 \subsubsection{Design Implication 5: Prioritize shared activities that could promote more communication.} As discussed above, grandparents and grandchildren always chose shared activities that foster the sharing of their experiences and life stories to deepen their understanding of each other better. This purpose is consistent with in-person meetings between them~\cite{kemp2005dimensions, moffatt2013connecting}. \rv{Our work also suggests three elements in VR that have the potential to promote more communication: familiar activities and environments, familiar avatar appearances, and easy-to-learn operations of VR controllers. 
 Future VR designers should leverage these elements to achieve better communication experiences between grandparents and grandchildren. }


 \subsubsection{Design Implication 6: Provide easy-to-learn and easy-to-use tools to support personal scene building.} We found that the virtual scenes that grandparents and grandchildren preferred to communicate in varied from pair to pair because of personal interests and family background. Therefore, we suggest that future VR applications should allow grandparents and grandchildren to customize scenes and activities to satisfy their shared interests and particular needs. Although some commercially available social VR applications only allow users to make minor modifications to existing scenes, such as changing the decoration and style of the dorm~\cite{recroomWebsite, mcveigh2019shaping}, or to customize their home in Unity and upload it to the VR world~\cite{VRChatHoneKit}, all such applications required users to have high technical sufficiency. Future VR designers and developers should design tools that allow novice grandparents and grandchildren to build their preferred virtual world scenes.

\sout{\textbf{Involving other members of the family if needed.} }
 \sout{Although previous work tried to reduce parental scaffolding and promote direct communication between grandparents and grandchildren by providing storytelling systems, digital books, or self-making devices~\cite{rodriguez2015sincom, wallbaum2018supporting, tibau2019familysong, forghani2018g2g}, participants in our study were willing to involve more family members when communicating in VR, especially the parents. In this case, grandparents and grandchildren can not only share their life stories and well-being with more family members but also enjoy the sense of reunion and happiness brought by family gatherings. Participants suggested that family meetings in VR should contain various activities and beautiful or meaningful scenery to foster conversation so that each member can participate and enjoy together. Prior work also showed that involving more family members in communication can generate more conversation topics and enhance the connection bond of families~\cite{kang2021momentmeld, forghani2018g2g, bucx2008intergenerational}. Current commercial VR applications also provide numerous activities that allow multiple users to enter the same virtual scene to interact, such as doing exercise, taking adventures, and watching movies~\cite{recroomWebsite, bigscreenWeb}. However, whether these activities are suitable for families and friendly for older adults is unknown.}
 \sout{The grandparent-grandchild connection is intertwined with the larger systems of family relationships \cite{mueller2003family}, so future VR applications should accommodate other family members if desired by the grandparents and grandchildren.}

\subsection{\rv{Developing Practical and Safe VR for Older Adults}}

 \rv{Our work uncovered three perceived barriers to adopting VR for intergenerational communication between grandparents and grandchildren: the overhead of setting up VR, the perceived difficulty of VR operations, and safety concerns. This finding is in line with prior work that shows older adults feel frustrated and insecure when using VR devices~\cite{heydari2020barriers, brown2019exploration, jcm9061986}. Caregivers play a critical role in supporting older adults in overcoming these barriers and enjoying the novel experiences brought by VR~\cite{waycott2022role}. However, for older adults who want to use VR independently, these barriers still negatively impact their experiences and prevent the adoption of VR. We propose the following design implications for creating practical and safe VR experiences for older adults in inter-generational communication.}

 \subsubsection{\rv{Design Implication 7: Provide an asymmetrical VR interaction that involves remote assistance from grandchildren.}} \rv{Previous work explored the design of asymmetric video games that leverage differences of technical proficiency between players to improve multiplayer engagement~\cite{harris2016leveraging}. 
 One possible asymmetrical design is to allow grandchildren to help their grandparents conduct the non-intuitive operations by gaining control of the grandparents' avatar and keeping the intuitive operations for grandparents to ensure they can enjoy the activity. Previous work showed that helping grandparents naturally occurs in everyday interactions~\cite{kennedy1992shared}. Our work also found that grandchildren always helped their grandparents carry out operations via voice instructions in VR. }
 The other asymmetrical design is to assign intuitive and straightforward operations to grandparents while allocating more complex operations to grandchildren. For instance, in a VR travel application, grandchildren can take a photo of their grandparents while the grandparents simply pose for the photo. The asymmetrical operations bring enjoyment by reducing technical requirements while fostering communication by increasing cooperation among grandparents and grandchildren. In this case, both parties can enjoy the fun of the skill-appropriate interactions.

 \subsubsection{\rv{Design Implication 8: Prevent falls and collisions to keep older adults safe during VR use.}} \rv{Participants worried about the risk of falls and collisions when older adults living alone use VR because the headsets make them blind from the physical world. We suggested that for older adults who live in residential care houses or with other family members, it would be better to get support from the caregivers when using VR~\cite{waycott2022role}. For older adults who want to use VR independently, future work should explore how to protect them from falls and collisions.
 Although previous work investigated falls occurring in general users when using VR (e.g., colliding with surroundings, and hitting spectators) and proposed possible solutions~\cite{dao2021bad, virk2006virtual}, the causes of falls in older adults are different when using VR due to the difference in mobility and cognitive abilities between older and younger people~\cite{ambrose2013risk}. Moreover, the consequences of falls in older adults tend to be more serious~\cite{ambrose2013risk}. Therefore, future work should explore what kinds of breakdowns would occur among older people when using VR and how to avoid these falls to keep older adults safe. For instance, providing a switch function for them to easily switch between the virtual world and their physical surroundings so that they can check whether there are any obstacles around them to make them feel safe. In this way, older adults can enjoy the benefits brought by VR while staying safe.}

\section{Limitation and future work}

 
 One limitation was that participants experienced VR for 30 minutes during the two-and-a-half-hour study. Thus, while participants were excited about VR in general, we could not rule out the possible novelty effect. As there was not enough time for participants to become familiar with VR usage, the difficulty of operating controllers was more prominent. Future work should consider a similar but longer-term study to better investigate how the increasing familiarity with VR might affect the perceptions of its usefulness among grandparents and grandchildren.
 
 Second, as our participants are all from an oriental cultural background, our results may not be directly generalizable to grandparents and grandchildren in other cultures. Future work should extend this work to investigate how different cultural backgrounds may affect the preferences of grandparents and grandchildren for VR-based remote inter-generational communication.

 \rv{Third, conducting a formal comparison study is important to understand grandparents' and grandchildren perceived advantages and drawbacks between VR, video-mediated communication, and other technologies. Future work is warranted to conduct a formal comparison to explore this difference in detail.}

 \rv{Fourthly, this work explored grandparents' and grandchildren's preferences and visions in VR-based remote communication to inform future design. However, what factors in VR (e.g., presence and immersion)  contribute to these preferences and how they work, are still unknown. Future work could explore these questions by performing a quantitative study.}

 \rv{Finally, according to our analysis, we found that the \textit{the technology acceptance of grandparents}, the \textit{closeness between grandparents and grandchildren}, and the \textit{health condition of the grandparent} might affect their willingness to use VR for remote communication. For example, the grandparents and grandchildren who had closer connections seemed to exhibit a higher acceptance of VR as their remote communication tool. Additionally, the \textit{gender of grandparents and grandchildren} might affect their preferences of shared activities in VR. For example, while grandfathers tended to share stories from their youth through VR activities, grandmothers preferred to have a family meeting or travel in VR to enjoy the happiness of chatting together with family members. Future work should explore what kind of VR activities are suitable for grandparents and grandchildren with different demographics to make inter-generational communication between grandparents and grandchildren more enjoyable.}

\st{we did not conduct a formal comparison study to examine participants' communication experiences in three different communication modes since they were already familiar with F2F and VC.  Future work is warranted to closely compare these types using counterbalanced study designs to validate the ratings in this study. }
 
\st{we explored participants' preferences for avatars, interactions, and shared activities, which may not be sufficient for the comprehensive design of VR for remote communication media for grandparents and grandchildren. Future work can extend this work in more aspects, such as taking the GUI design of the VR system into account.}

\section{Conclusion}

A close inter-generational connection between grandparents and grandchildren is beneficial for both generations. However, maintaining a close inter-generational connection is challenging due to the increasingly dispersed. \rv{We sought to understand how VR might be leveraged to facilitate inter-generational communication between grandparents and grandchildren. Through a user-centered participatory design study with twelve pairs of grandparents and grandchildren, we uncovered participants' preferences and visions about three key elements of VR to foster better inter-generational communication: \textit{avatar appearance}, \textit{interpersonal interaction}, and \textit{shared activities}. We also revealed \textit{their perceived barriers to adopting VR as communication media}.
Based on our findings, we further derived eight design implications and discussed future directions for VR designers and researchers to consider when designing VR to support remote communication between grandparents and grandchildren.}
\st{We investigated the feasibility of using VR for remote communications between grandparents and grandchildren.
Through a mixed-methods user study with twelve pairs of grandparents and grandchildren, we identified the pros and cons of VR in facilitating remote communication and explored their preferences of avatar appearance and shared activities on VR-based remote inter-generational communication. Our participants regarded VR as \textit{communication facilitator} and \textit{nostalgia reliever}. We revealed the uniqueness of VR communication in grandparent-grandchild relationships compared to other relationships. Finally, we proposed five design implications derived from our results, which inform future VR-based remote communication between grandparents and grandchildren. }



\bibliographystyle{ACM-Reference-Format}
\bibliography{main}

\end{document}